\documentclass[aps,prb,twocolumn,groupedaddress,showpacs]{revtex4-1}

\usepackage[T1]{fontenc}
\usepackage[pass]{geometry}
\usepackage{lipsum}
\usepackage{mathrsfs}
\usepackage{epsfig}
\usepackage{amsmath}
\usepackage{color}
\usepackage{subfigure}
\usepackage{float}

\DeclareMathOperator{\sech}{sech}

\DeclareMathOperator{\arctanh}{arctanh}

\DeclareMathOperator{\arccoth}{arccoth}

\newcommand{\pvec}[1]{\vec{#1}\mkern2mu\vphantom{#1}}

\newcommand{\be}{\begin{equation}}
\newcommand{\ee}{\end{equation}}
\newcommand{\bea}{\begin{eqnarray}}
\newcommand{\eea}{\end{eqnarray}}

\begin{document}

\title{Analytic theory for the switch from Bloch to N\'eel domain wall in nanowires with perpendicular anisotropy}

\author{M.~D.~DeJong}
\author{K.~L.~Livesey}
\email{klivesey@uccs.edu}

\affiliation{Center for Magnetism and Magnetic Materials, Department of Physics and Energy Science, University of Colorado Colorado Springs,1420 Austin Bluffs Pkwy, Colorado Springs, CO 80918, USA}

\date{\today}
\begin{abstract}
Rectangular nanowires may have domain walls nucleated and moved through them to realize many devices. It has been shown that at a particular width and thickness of a nanowire with perpendicular anisotropy, there is a switch from the domain wall being of Bloch-type to it being of N\'eel-type. This critical shape can be found through micromagnetics simulations, but here we present an analytic calculation for the energies of both wall types involving two iterations for the demagnetizing energy density. The expressions developed are long, but have the advantage that by simply inputting material parameters for the magnetic material, the critical shape can be found using a calculator in a matter of seconds. We compare our results to those found using micromagnetics and those found experimentally and the agreement is good.
\end{abstract}
\pacs{75.30.-m, 75.60.Ch, 75.75.Cd, 85.75.-d}

\maketitle


\section{Introduction}

In the last few years, the ability to tailor the shape, width, speed and chirality of magnetic domain walls in nano-materials has been demonstrated using interface and surface effects. Such effects do not play a significant role in bulk but lead to dramatic changes in nano-objects. For example, the Dzyaloshinskii-Moriya interaction (symmetry-allowed at surfaces and interfaces) has been shown to enhance one domain wall chirality over another in thin films and rectangular nanowires. \cite{Chen,Boulle,Franken,Chen2,Martinez2} Another example is the ability to alter the strain in a material and therefore the magnetic anisotropy constant and the domain wall velocity, with an applied voltage. \cite{Chiba,Lei,Shepley} This is due to interfacial coupling between the magnetic thin film or nanowire and a piezoelectric material. Nanometer-sized domain walls are being manipulated in ways that could not be realized just a few years ago.

Domain walls in magnetic nanowires have also been proposed recently for a variety of important applications \cite{Sampaio} including domain wall logic \cite{Allwood}, miniaturized information storage \cite{Parkin} and bio-sensing.\cite{Rapoport} Materials with perpendicular magnetic anisotropy (PMA) are often preferred over materials where the magnetization in domains lies in-plane -- along the nanowire's long axis -- because the domain walls are narrower. \cite{Fukami} 

\begin{figure}[b]
\begin{center}
\includegraphics[width=8.5cm]{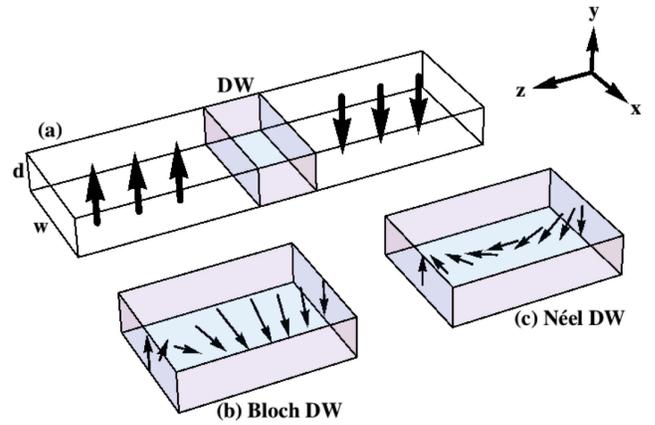}
\caption{\label{geometry}(Color online) (a) The geometry of a rectangular nanowire with perpendicular magnetic anisotropy, width $w$ and thickness $d$. Arrows show the direction of magnetization. Depending on the size of $w$ and $d$, the lowest energy domain wall may be of (b) Bloch or (c) N\'eel type. In the Bloch wall, the magnetization rotates perpendicular to the wire's long $z$ axis, in the $x$-$y$ plane. In the N\'eel wall, the magnetization rotates along the wire's long $z$ axis, in the $y$-$z$ plane.}
\end{center}
\end{figure}

For many of these applications it is desirable to \emph{move} a domain wall, for example, past a read head. It has been shown that this can be done with the least amount of energy when the aspect ratio of a rectangular nanowire is just right so as to see a switch between the equilibrium domain wall configuration being of N\'eel-type to it being of Bloch-type. \cite{Jung,Martinez,Koyama,Demiray} A schematic diagram of a Bloch and N\'eel wall in a nanowire with PMA is shown in Fig.~\ref{geometry}, with the nanowire's length along the $z$ axis. The width of the wire is $w$ and the thickness is $d$. 

Such a transition from Bloch to N\'eel wall has been know to occur in thin films since the 1950s, as a function of the film thickness. \cite{Neel1955,Middelhoek} The transition thickness can be found analytically for thin films. \cite{Aharoni} Some work has also been done on studying a similar Bloch-N\'eel transition in wide strips with in-plane magnetization where the domain wall occurs along the width (our $x$) direction. \cite{Rivkin} For our geometry (nanowires with PMA) the switch from one type to the other occurs because as the nanowire is made narrower ($w \to 0$), the surface demagnetizing energy of the Bloch wall due to surfaces in the $x$ direction increases, while the volume demagnetizing energy of the N\'eel wall decreases. Hence there is a transition from Bloch walls at larger widths to N\'eel walls at smaller widths. There are other competing energy contributions (anisotropy, exchange and demagnetizing energy due to the $y$ -- top and bottom -- surfaces of the nanowire) but the demagnetizing energy change described here is the essential reason for the transition.

In this work, we calculate using a one dimensional model the aspect ratio where the switch between a Bloch and N\'eel wall having the lowest energy occurs in nanowires with PMA. The calculation for the respective energies of the two walls is \emph{analytic} and this represents the major advantage of our results over those found using micromagnetics: by plotting the domain wall energies, the aspect ratio where the switch occurs can be found in a matter of seconds. We hope therefore that our results will be useful to experimentalists looking to design and make nanowires for domain wall applications. 

The major issue we face is calculating the demagnetizing energy of the domain walls analytically. A two-step calculation to approximate this energy contribution is utilized and comparisons to numerical and literature results indicate that this method is robust. The calculation will be presented in two stages in Section~\ref{calc}. Our results along with comparison to results from the literature will be shown in Section~\ref{results}. Finally, conclusions are given in Section~\ref{conclusion}. As a note to the reader, throughout this paper the terms nanowire and strip will be used interchangeably.


\section{Calculation}
\label{calc}

In order to obtain an analytic solution for the energy of the Bloch and N\'eel walls in a rectangular nanowire, an approximation for the demagnetization energy must be used. This is because the magnetization configuration (ie. the wall shape) alters the demagnetizing energy, which in turn alters the magnetization configuration. In other words, we have a self-consistent problem.

Our approximation consists of two steps. Firstly, we use demagnetizing factors for a uniformly magnetized strip to estimate the domain wall shape and width. Secondly, we use this solution to solve analytically for the demagnetizing energy of the strip with a Bloch or N\'eel wall present. Although these two iteration steps may appear to be a crude approximation, the theory agrees exceptionally well with micromagnetic and experimental results found elsewhere. Moreover, the domain wall width that can be found numerically does not markedly differ from that found in our analytic approximation.

In this section we will describe our methods by dividing the two steps into separate subsections. In a third subsection we will then compare the results to those of a semi-numerical method to underline the validity of the two-step approximation. In the semi-numerical method, the domain wall width is found using energy minimization, rather than using an analytic expression. 


\subsection{First iteration}
\label{first}

For a \emph{uniformly} magnetized strip with infinite length in the $z$ direction, finite width $w$ in the $x$ direction and finite thickness $d<w$ in the $y$ direction (as illustrated in Fig.~\ref{geometry}), the demagnetizing energy density is due to surface terms and is given in SI units by
\be
\mathscr{E}_{d} = \frac{1}{2} \mu_{0} \left(  N_{x} M_{x}^2 + N_{y} M_{y}^{2}   \right), 
\label{uniformdemag}
\ee
where $\vec{M}$ is the magnetization vector, $\mu_{0}$ is the permeability of free space, and $N_{x}$ and $N_{y}$ are the demagnetizing factors with $N_{x}+N_{y}=1$. These factors have been found by Brown \cite{Brown} and are given by
\be
N_{y} = \frac{2}{\pi} \arctan \left(\frac{1}{p}\right) + \left(\frac{1-p^{2}}{2 \pi p}\right)\ln(1+p^{2}) + \frac{p}{\pi}\ln(p),
\label{demagfactor}
\ee
where $p=d/w$ is the aspect ratio of the nanowire's rectangular cross-section, as illustrated in Fig.~\ref{geometry}. Eq.~\eqref{uniformdemag} is an approximation as the dipolar fields vary throughout the strip's cross-section, rather than being constant. The demagnetizing factors of Brown (Eq.~\eqref{demagfactor}) take the average of the dipolar fields through the width and thickness of the strip. At times it may also be appropriate to use the value of the dipolar field at the strip's center to find approximate demagnetizing factors. \cite{Brown}

Even though Eq.~\eqref{uniformdemag} is an approximation only valid for uniform magnetization, here we assume that the magnetization varies along the nanowire, in the $z$ direction only. Two angles define the magnetization direction: $\theta(z)$ is the angle of magnetization in the $y$-$z$ plane with $\theta=0$ corresponding to magnetization aligned with the $z$ axis, and $\phi(z)$ is angle of the magnetization in the $x$-$y$ plane with $\phi=0$ for $\vec{M}$ aligned with the $x$ axis. In other words, $\theta$ is the relevant angle for the N\'eel wall and $\phi$ is the relevant angle for the Bloch wall. In this way, solving for the domain wall configuration reduces to a one dimensional problem. We can re-write Eq.~\eqref{uniformdemag} using these angles:
\be
\mathscr{E}_{d}(z) =
\label{demag2}
 \frac{1}{2} \mu_{0} M_{s}^2
\left\{
\begin{array}{ll}
 N_{y} \sin^2 \theta(z)   , & \textrm{N\'eel} \\
 N_{x} \cos^2 \phi(z) + N_{y} \sin^2 \phi(z)   , & \textrm{Bloch}  
\end{array}
\right. .
\ee
This equation is clearly non-sensical as the demagnetizing factors are only valid for uniform magnetization but this will serve as a first approximation for the demagnetizing energy. In particular, the volume contribution to the dipolar energy is clearly not included in this first iteration, since the demagnetizing factors only account for surface contributions.

The other energy density contributions to the nanowire include exchange and anisotropy. The exchange energy density is given by \cite{Hubert} 
\be
\mathscr{E}_{ex}(z) =A\left[\sin^{2}\theta ~\left(\frac{d\phi}{dz}\right)^{2} +\left(\frac{d\theta}{dz}\right)^{2} \right],
\label{exchange}
\ee
where $A$ is the exchange constant with units of J/m. The perpendicular anisotropy contributes an energy density
\be
\mathscr{E}_{a}(z) = K \cos^{2}\phi(z),
\label{anisotropy}
\ee
for the Bloch wall where $K$ is the anisotropy constant with units of J/m$^{3}$. For the N\'eel wall $\phi$ is replaced with $\theta$ in Eq.~\eqref{anisotropy}. This expression states that the anisotropy energy is minimized if the magnetization is aligned with either of the $\pm y$ axes, as illustrated in Fig.~\ref{geometry} in the two domain regions.

The energy per unit area $E$ of the strip is given by integrating the total energy density along the $z$ direction, namely
\be
E= \int_{-\infty}^{\infty} \mathscr{E} \left( \theta, \phi, \dot{\theta}, \dot{\phi} \right) dz,
\label{EperunitA}
\ee
where 
\be
\mathscr{E} \left( \theta, \phi, \dot{\theta}, \dot{\phi} \right) = \mathscr{E}_{d} + \mathscr{E}_{ex} + \mathscr{E}_{a}.
\label{Eintegrand}
\ee
Here the dots over the angles signify the derivative is taken with respect to position $z$. Note that for the N\'eel wall there is only dependence on $\theta$ and its derivative, while for the Bloch wall there is only dependence on $\phi$ and its derivative. 

We find the domain wall configuration using calculus of variations to minimize the energy in Eq.~\eqref{EperunitA} via the well-known Euler-Lagrange equation. \cite{Coey} For the Bloch wall, for example, we look for a solution to the equation 
\be
\frac{\delta \mathscr{E} }{ \delta \phi } - \frac{d}{dz} \frac{ \delta \mathscr{E} }{ \delta \dot{\phi} } = 0,
\ee
where $\delta/\delta \phi$ signifies a functional derivative and the equation is subject to the boundary conditions $\phi(\pm \infty)=\pm \frac{\pi}{2}$ and $\dot{\phi}(\pm \infty) =0$. 

The magnetization angle for the Bloch wall is found to be 
\be
\phi(z) =2 \arctan \left(\frac{e^{z/L_{B}}-1}{e^{z/L_{B}}+1} \right),
\label{magnetizationangle}
\ee
with $L_{B}$ the characteristic domain wall width. The normalized magnetization components $M_{x}(z)/M_{s}=\cos \phi(z)$ and $M_{y}(z)/M_s = \sin \phi(z)$ are found to be
\begin{eqnarray}
\frac{M_{x}(z)}{M_{s}} &=& \sech \left({\frac{z}{L_{B}}} \right),
\label{xcomp} \\
\frac{M_{y}(z)}{M_{s}} &=& \tanh \left({\frac{z}{L_{B}}} \right).
\label{ycomp}
\end{eqnarray}
They are shown by solid lines in Fig.~\ref{magapprox} (panels (a) and (b) respectively) and follow expected behavior for Bloch wall rotation. 

The characteristic width $L_{B}$ for the Bloch wall is 
\be
L_{B} =\sqrt{\frac{A}{K-\frac{1}{2} \mu_{0} M_{s}^{2} (2N_{y}-1)}},
\label{wallwidth}
\ee
where $M_{s}$ is the saturation magnetization with $M_{s}^{2} =M_{x}^{2}+M_{y}^{2}+M_{z}^{2}$. This is the well known result that the domain wall width is given by the square root of the exchange constant divided by the effective anisotropy constant. \cite{Coey} Here, the demagnetization fields in the $x$ and $y$ directions act to either reduce or increase the anisotropy field along the $y$ direction, depending on the size of $N_{y}$. In turn, the size of $N_{y}$ depends purely on the aspect ratio of the rectangular nanowire. Therefore, we see already in this first iteration that the aspect ratio strongly affects the wall width and therefore the energy of the wall.

For the N\'eel wall, the same Eq.~\eqref{magnetizationangle} holds, but with $\theta$ replacing $\phi$ and the characteristic N\'eel wall width, $L_{N}$ replacing $L_{B}$. $L_{N}$ is given by
\be
L_{N} =\sqrt{\frac{A}{K-\frac{1}{2} \mu_{0} M_{s}^{2} N_{y}}}.
\label{wallwidthneel}
\ee
Again one sees the effective anisotropy is altered by demagnetizing effects. For the N\'eel wall, the finite size of the nanowire always acts to reduce the effective anisotropy and therefore increase the wall width. For thinner wires ($d\downarrow$) or, alternatively, for wider wires ($w\uparrow$), $N_{y}$ gets larger and the wall width increases. In the thin film limit ($w\to\infty$, $N_{y}\to1$) $L_{N}=L_{B}$.

Equation~\eqref{magnetizationangle} is substituted back into the energy per unit area Eq.~\eqref{EperunitA} and then Eq.~\eqref{EperunitA} can be integrated to find the total energy of the walls. Explicitly, the demagnetizing energy per unit area for the Bloch wall (calculated by integrating between $z=-D$ and $z=D$ where $z=0$ is the center of the wall) becomes
\begin{eqnarray}
E_{dB}(D) =& &-\mu_{0}M_{s}^{2} L_{B} \left[ N_{x} \tanh \left({\frac{D}{L_{B}}}\right) \right. \nonumber \\
	& & \left. + N_{y}\left(\frac{D}{L_{B}} - \tanh \left({\frac{D}{L_{B}}}\right) \right)     \right],
\label{demagbloch}
\end{eqnarray}
The subscript $B$ in $E_{dB}$ indicates that this energy applies to the Bloch wall but the same process of substitution and integration can be used to express the N\'eel wall energies using the appropriate forms of Eqs.~\eqref{EperunitA}~and~\eqref{magnetizationangle}. Notice that Eq.~\eqref{demagbloch} diverges as $D \to \infty$. This is artificial and is due to defining the domains at $z\to \pm \infty$ to have non-zero energy density. By subtracting the domain energy density ($\mathscr{E}_\textrm{dom} = \frac{1}{2} \mu_{0} N_{y} M_{s}^{2} $), the energy no longer diverges.

The exchange energy per unit area for the Bloch wall is given by 
\be
E_{exB} (D) = \frac{2A}{L_{B}} \tanh \left({\frac{D}{L_{B}}}\right),
\label{exchangebloch}
\ee
and the anisotropy energy per unit area is given by
\be
E_{aB} (D) = 2KL_{B} \tanh \left({\frac{D}{L_{B}}}\right).
\label{anisotropybloch}
\ee
Notice that the exchange energy decreases for a longer wall while the anisotropy energy contribution goes up. Eqs.~\eqref{exchangebloch}~and~\eqref{anisotropybloch} are used to calculate the respective energy contributions in this iteration and in the next. However, the demagnetizing energy Eq.~\eqref{demagbloch} requires a more accurate description.

Using the crude approximation that the dipolar energy can be accounted for using demagnetizing factors and surface effects alone, we find that the N\'eel wall always has lower energy than the Bloch wall. The N\'eel wall occupies a longer region ($L_{N }~>~L_{B}$) meaning that its anisotropy energy is increased compared to the Bloch wall, but demagnetizing and exchange contributions for the N\'eel wall decrease by a larger amount which accounts for the lower energy state of the N\'eel wall for all aspect ratios. Obviously there is no energy switch between Bloch and N\'eel walls so we move on to a second iteration for the demagnetizing energy. 


\subsection{Second iteration}
\label{second}

To obtain a more accurate comparison of the demagnetizing energy for Bloch and N\'eel walls, the magnetization angles $\phi(z)$ and $\theta(z)$ in Eq.~\eqref{magnetizationangle} resulting from the first iteration are used to calculate the magnetostatic potential. This is then used to find the dipolar fields and the total demagnetizing energy density at any position $\vec{r}$. The total magnetostatic potential -- volume and surface contributions together -- at position $\vec{r}=(x,y,z)$ is given by \cite{Aharoni,Jackson}
\be
U(\vec{r})= \frac{1}{4\pi} \int_{V'} \frac{\vec{M}(\pvec{r}') \cdot (\pvec{r}-\pvec{r}' ) }{{|\pvec{r}-\pvec{r}' |}^{3}}d\pvec{r}'
\label{potential}
\ee
where the primed coordinates denote the source of magnetization in the material and the integral is performed over the volume of the nanowire. The solution for $\vec{M}(\pvec{r}') \equiv \vec{M}(z')$ from Section~\ref{first} -- for the Bloch wall, this is given by Eqs.~\eqref{xcomp} and \eqref{ycomp} -- is substituted into Eq.~\eqref{potential}. However, this  produces a nonintegrable expression. 
 
\begin{figure}[b]
\begin{center}
\includegraphics[width=8.5cm]{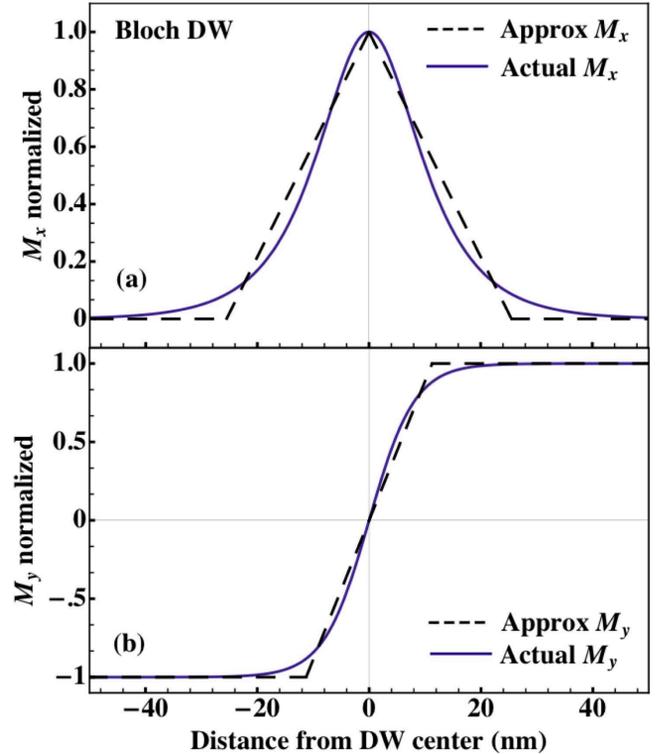}
\caption{\label{magapprox}(Color online) Linear approximations (dashed lines) compared to the magnetization components (solid lines, Eqs.~\eqref{xcomp} and \eqref{ycomp}) obtained from the first iteration for the Bloch wall, both as a function of $z$. Panel (a) shows the in-plane component $M_{x}(z)$ and panel (b) shows the out-of-plane component $M_{y}(z)$ of the magnetization, consistent with Figure~\ref{geometry}(b). The material parameters are $M_{s}=3\times10^{5}$~A/m, $A=10\times10^{-12}$~J/m, $K=2\times10^{5}$~J/m$^{3}$, $d=3$~nm and $w=60$~nm. Similar linear approximations are used for the $M_{y}(z)$ and $M_{z}(z)$ components for the N\'eel wall. } 
\end{center}
\end{figure}
 
To bypass this nonintegrable equation, piecewise linear approximations for the magnetization components are utilized and replace the hyperbolic secant and hyperbolic tangent expressions in Eqs.~\eqref{xcomp}~and~\eqref{ycomp}, respectively. The approximations are illustrated by dashed lines in Fig.~\ref{magapprox}. The piecewise approximation used for $M_{x}$ for the Bloch wall is given by
\begin{equation}
\label{piecewise}
  \frac{M_{xB} (z)}{M_s} \sim
    \left\{
      \begin{array}{@{}l  r @{} c @{} l}
        0, & -\infty {}<{} & z & {}< -b_{xB} \\
        1+\frac{z}{b_{xB}}, & -b_{xB} \leq{} & z & {}< 0 \\
        1-\frac{z}{b_{xB}}, & 0 \leq{} & z & {}\leq b_{xB} \\
        0, & b_{xB} <{} & z &  {}<{} \infty \\
      \end{array}
    \right. ,
\end{equation}
with $z$ the distance from the center of the domain wall, as illustrated in Fig.~\ref{magapprox}, and the length $b_{xB}$ an adjustable fit parameter. The same process is used to approximate the $M_{y}$ component for the Bloch wall
\begin{equation}
\label{piecewise2}
  \frac{M_{yB} (z)}{M_s} \sim
    \left\{
      \begin{array}{@{}l  r @{} c @{} l}
        -1, & -\infty {}<{} & z & {}<{} -b_{yB} \\
        \frac{z}{b_{yB}}, & -b_{yB} \leq{} & z & {}\leq b_{yB} \\
        1, & b_{yB} {}<{} & z & {}<{}\infty \\
      \end{array}
    \right. ,
\end{equation}
and both the $M_{y}$ and $M_{z}$ components for the N\'eel wall, with different resulting adjustable fit lengths. The adjustable fit length varies for each wall type, each component, and with each combination of nanowire dimensions and material parameters. To obtain the best fit for each possible set, the area under the piecewise approximation (dashed lines in Fig.~\ref{magapprox}) and the magnetization components (solid lines in Fig.~\ref{magapprox}) is equated. For the $M_{x}$ component of the Bloch wall this means that
\be
\int_{-\infty}^{\infty} \frac{M_{xB}(z)}{ M_{s} } dz = \int_{-\infty}^{\infty} \cos \phi(z) dz.
\label{equatearea}
\ee
Equation~\eqref{equatearea} is then solved directly for the adjustable length $b_{xB}$ and the result can be expressed in terms of the characteristic wall widths given in Eqs.~\eqref{wallwidth} and \eqref{wallwidthneel}. We find that $b_{xB}=\pi L_{B}$ obtains the best fit for the $x$ component of the Bloch wall. For the $y$ component of the Bloch wall, the length is found to be $b_{yB}=\ln(4) L_{B}$. For the N\'eel wall, the fitted lengths are given by $b_{yN}=\ln(4) L_{N}$ and $b_{zN}=\pi L_{N}$ respectively.

Utilizing the linear approximations for the magnetization components allows for calculation of the potential via Eq.~\eqref{potential} and then the auxiliary dipolar field 
\be
\vec{H}(\vec{r}) =-\nabla{U(\vec{r})}, 
\label{strayfield}
\ee
which has three dimensional dependence. In accordance with the one dimensional model applied, the components of $\vec{H}(\vec{r})$ must be approximated to have one dimensional dependence, ie. $\vec{H}(\vec{r}) \sim \vec{H}(z)$. To do this, two methods are applied. The  component $H_{x}$ that is non-zero only for the Bloch wall is averaged over the width ($-w/2<x<w/2$) and thickness ($-d/2<y<d/2$) of the strip. For the other two components $H_{y}$ and $H_{z}$, their values in the center of the strip ($x=y=0$) are taken as being representative of the average. This reduces the dimensionality of the problem from three to one and allows us to progress in calculating the demagnetizing energy density analytically. The resulting analytic expressions for $H_{x}(z)$, $H_{y}(z)$, and $H_{z}(z)$ are lengthy and thus included in the Appendix as Eqs.~\eqref{Hxappendix}-\eqref{Hzappendix}.

\begin{figure}[H]
\begin{center}
\includegraphics[width=8.5cm]{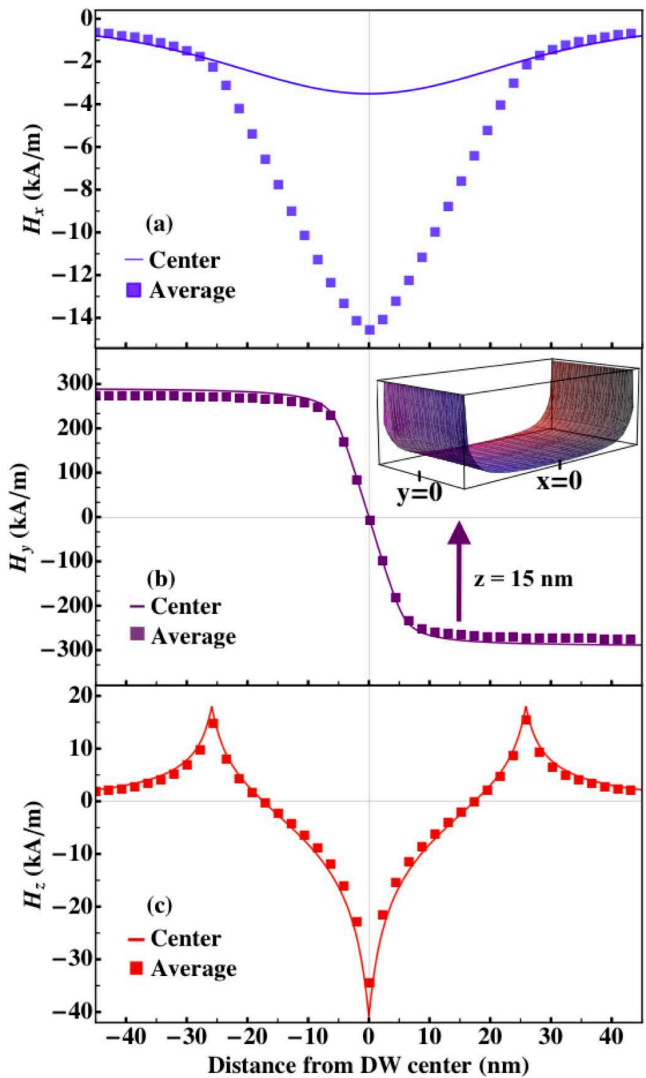}
\caption{\label{Hfield}(Color online) Plots of approximations for the auxiliary field components (a)~$H_{x}$, (b)~$H_{y}$, and (c)~$H_{z}$, as a function of distance $z$ along the wire. Values are calculated by either averaging over the width and thickness of the nanowire (squares) or by evaluating the field at the center of the nanowire where $x=y=0$ (solid lines).  Material parameters used to generate these plots are $w=60$~nm, $d=3$~nm, $M_{s}=3\times10^{5}$~A/m, $A=10\times10^{-12}$~J/m, and $K=2\times10^{5}$~J/m$^{3}$, and with $L_{B}$ used to generate panels (a) and (b) and $L_{N}$ for panel (c). $H_{y}$ and $H_{z}$ show reasonable agreement between the average field and the value at the center of the strip justifying the choice of these components at the center of the strip in calculating the demagnetizing energy. $H_{x}$ must be analytically averaged to maintain the appropriate shape of the auxiliary field. The inset in (b) shows how $H_{y}$ varies with width ($x$ direction) and thickness ($y$ direction) at a position $z=15$~nm from the domain wall center.}
\end{center}
\end{figure}

The reason for averaging or taking the value of the field at the strip center is explained with the aid of Fig.~\ref{Hfield}. Figure~\ref{Hfield} illustrates components of $\vec{H}(x,y,z)$ that are either (i) numerically averaged over the wire cross-sectional area (squares) or are (ii) taken as the values at the center of the strip $\vec{H}(0,0,z)$ (solid lines). Panel (a) shows the result for $H_{x}$, panel (b) for $H_{y}$ and panel (c) for $H_{z}$, where $H_{x}$ and $H_{z}$ were generated from Bloch and N\'eel parameters, respectively, and $H_{y}$ is plotted here for the Bloch wall but could also easily be plotted for the N\'eel wall by substituting in the appropriate characteristic wall width $L_{N}$. The $H_{x}$ numerically averaged field differs markedly from the value of the field at the center of the strip because $H_{x}$ has maximum amplitude at the edge of the strip. In contrast, $H_{y}$ and $H_{z}$ are essentially flat functions with a drop in amplitude near the edges in the $x$ direction, so the average value is similar to the value in the center. To show this, an example of the $H_{y}(x,y,z)$ component as a function of $x$ and $y$ is shown in the inset of Fig.~\ref{Hfield}(b) with $z=15$~nm and it can be seen that the function is essentially flat apart from $H_{y} \to 0$ as $x\to \pm w/2$. Notice in panel (c) that $H_{z}$ has sharp peaks around 26~nm from the center of the domain wall. This corresponds to where $M_{z}(z)$ has a discontinuity due to the linear approximations, illustrated in Fig.~\ref{magapprox} for the $M_{x}$ and $M_{y}$ components. 

The magnitude of $H_{y}$ can be checked using the expected value far from the domain wall region. Far into the domains, the auxiliary field is that of a uniformly magnetization strip with $H_{y}=-N_{y}M_{y}=-N_{y} M_{s}$. For example, for the parameters used in Fig.~\ref{Hfield}, $N_{y}=0.93$ using Eq.~\eqref{demagfactor} and $M_{s}=3\times10^{5}$~A/m, resulting in a field with strength $H_{y}\sim279$~kA/m, in agreement with the edges of Fig.~\ref{Hfield}(b). Note also that the averaging was checked by comparing to the exact numerical solution of 
Eqs.~\eqref{potential} and \eqref{strayfield} without the linear approximations for $\vec{M}$ and the numbers agree well. 

The components of the auxiliary field are then used to calculate the demagnetizing energy density for Bloch and N\'eel walls. The crude form of  $\mathscr{E}_{d}$ in Eqs.~(\ref{uniformdemag}) and (\ref{demag2}) which utilizes demagnetizing factors is replaced by
\be
\mathscr{E}_{d}(z) =-\frac{1}{2}\mu_{0}   \vec{M} (z) \cdot \vec{H}(z).
\label{demagenergy}
\ee
Here the $x$ and $y$ components of $\vec{M}$ and $\vec{H}$ are nonzero for the Bloch wall and the $y$ and $z$ components are nonzero for the N\'eel wall. Substituting the linear magnetization approximations in Eqs.~\eqref{piecewise} and \eqref{piecewise2} (and the appropriate forms of these for the N\'eel wall) and the auxiliary field components in Eqs.~\eqref{Hxappendix}-\eqref{Hzappendix} into Eq.~\eqref{demagenergy}, then integrating between $z=\pm D$, allows calculation of the demagnetizing energy per unit area of the two wall types of the form
\be
E_{d}(D) = \frac{1}{2} \mu_{0} M_{s}^{2} \left[ N_{xx}(D)+N_{yy}(D)+N_{zz}(D) \right],
\label{demagenergy2}
\ee
where the $N_{ii}$ factors are given in the Appendix in Eqs.~\eqref{Nxx}-\eqref{Nzz} and have units of length. Note that Eq.~\eqref{demagenergy2} still represents an approximation since the magnetization profile used firstly assumes uniform demagnetizing factors and secondly has been linearized. However, it is a better approximation than in the first iteration and we will show in Section~\ref{results} that it works exceptionally well at finding the dimensions of the nanowire which correspond to the Bloch-N\'eel energy crossing.

This analytic demagnetizing energy per unit area Eq.~\eqref{demagenergy2} is then combined with the exchange energy per unit area, Eq.~\eqref{exchangebloch}, and anisotropy energy per unit area, Eq.~\eqref{anisotropybloch}. Then the energies of the two wall types are calculated for given material parameters $K$, $A$ and $M_{s}$, for a given integration range $D$, and for a given geometry $d$ and $w$. It is found that there is a particular aspect ratio $p=\frac{d}{w}$ of a nanowire in which the Bloch and N\'eel wall types exhibit equal energy per unit area, as is predicted through micromagnetic simulations \cite{Martinez} and measured in experiment. \cite{Koyama} Direct comparison of our analytic findings to experimental and micromagnetic works will be made in Section~\ref{results}. Note that as long as the integration range $D$ is chosen so as to cover the full domain wall width, the results are insensitive to the range that is chosen.


\subsection{A check via energy minimization}
\label{numericcomparison}

Before moving to Section~\ref{results}, a check of the two-step approximation for the demagnetizing energy is presented here. Remember that solving for the domain wall solutions and their demagnetizing energies is in fact a self-consistent problem so the question is: does using just two iterations approximate the domain wall solution well? Instead of calculating a domain wall width $L_{B}$ or $L_{N}$ using demagnetizing factors and then using these to plot the energy per unit area, we can instead leave the domain wall width as a free parameter and numerically minimize the analytic energy per unit area to find it. This is similar to the approach outlined by Aharoni, \cite{Aharoni} where a thin film with in-plane magnetization was treated and it was found that there is a critical film thickness $d$ for which the equilibrium domain wall type switches from Bloch to N\'eel.

\begin{figure}[b]
\begin{center}
\includegraphics[width=8.5cm]{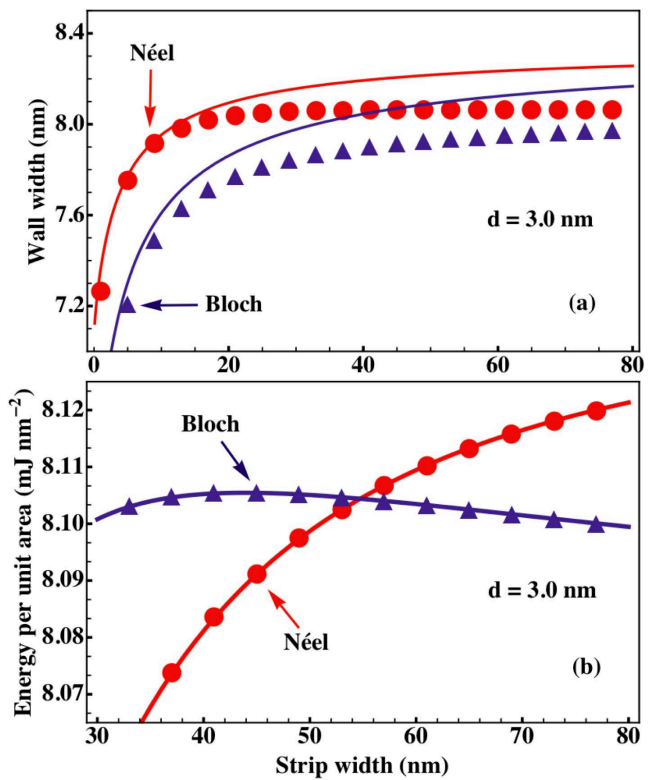}
\caption{\label{numericcompare}(Color online) Panel (a) shows comparison of Bloch and N\'eel domain wall widths calculated analytically (solid lines) and through minimizing the total energy per unit area with respect to wall width (shapes), following Aharoni in Ref.~\cite{Aharoni}. Parameters used are $d=3$~nm, $D=30$~nm, $M_{s}$~=~$3\times10^{5}$~A/m, $A$~=~$10\times10^{-12}$~J/m, and $K=2\times10^{5}$~J/m$^{3}$. Both methods yield similar wall widths and produce negligible energy differences, as shown in (b), strengthening the analytic results.} 
\end{center}
\end{figure}
 
The energy density contributions -- exchange, anisotropy and demagnetizing, given in Eqs.~\eqref{exchange}, \eqref{anisotropy} and \eqref{demagenergy} respectively -- are all the same but now have dependence on both position $z$ along the strip and also on the Bloch/N\'eel wall widths $L_{B}'/L_{N}'$. Here the primes represent that these are different widths from those analytically calculated earlier, as we are keeping the widths as free parameters here. Note that the linear approximation of the magnetization profile is used for the demagnetizing energy but the arctangent solution Eq.~\eqref{magnetizationangle} is used to find the other two energy density contributions, as was the case with our calculations in Section~\ref{second}. By integrating the total energy density on either side of the center of the Bloch wall, the energy per unit area for the Bloch wall is calculated
\be
E(L_{B}') = \int_{-D }^{D} \mathscr{E}(z,L_{B}') dz.
\ee
The energy per unit area is an analytic expression that can be numerically minimized to find the equilibrium value of $L_{B}'$ and therefore the Bloch wall's minimum energy. The same can be done for the N\'eel wall.
 
Figure~\ref{numericcompare}(a) compares the analytically calculated wall widths, Eqs.~\eqref{wallwidth}~and~\eqref{wallwidthneel} (solid lines) with the numerically calculated wall widths using energy minimization described in this subsection (shapes) for a particular set of material parameters indicated in the caption. The wall widths are shown for both Bloch (triangles) and N\'eel (circles) wall types and the trends using both methods are in agreement. The wall widths calculated using the two different methods agree best at small strip widths. At large strip widths, the difference between the two methods is less than 3\% for both wall types. 

Furthermore, when the wall widths from either method are inserted back into the energy per unit area, Eq.~\eqref{EperunitA}, and the total energy per unit area is calculated, the difference in energy between the two methods is negligible. This is shown in Fig.~\ref{numericcompare}(b) where the total energy per unit area (with integration range $D=30$~nm) is plotted as a function of the strip width. The results using energy minimization (shapes) exactly overlay the analytic results (solid lines) and this was checked against several sets of material parameters and several values of $D$. Notice in Fig.~\ref{numericcompare}(b) that there is a switch between the Bloch and N\'eel wall being lower in energy that occurs for a width $w=55$~nm. This is for a fixed thickness $d=3$~nm. The reason for the switch will be discussed in more detail in the next section.

This comparison to an alternate solution method strengthens the analytic results presented here. The advantage of using the analytic expressions for the domain wall characteristic widths, rather than those found numerically using energy minimization, is that one may calculate the energy of both domain wall types quickly and easily knowing just the geometry and material parameters of the nanowire. It should be noted that the energy minimization method discussed in this subsection assumes that the domain wall shape is as described in Eq.~\eqref{magnetizationangle} and only its characteristic width can vary. Assuming an alternate functional form for the domain wall may alter the results.


\section{Results and discussion}
\label{results}
\begin{figure}[b]
\begin{center}
\includegraphics[width=8.5cm]{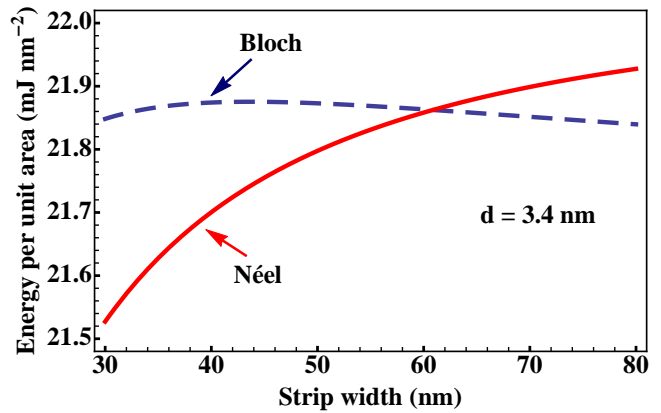}
\caption{\label{energyswitch}(Color online) The energy per unit area of Bloch and N\'eel walls as a function of nanowire or strip width. As the width of the nanowire is increased, the Bloch domain wall becomes energetically favorable over the N\'eel domain wall. Here $d= 3.4$~nm,  $M_{s} = 6.6\times10^{5}$~A/m, $A=15\times10^{-12}$ J/m, and $K=4.1\times10^{5}$ J/m$^{3}$, and the integration range is between $z=\pm30$ nm. The critical width where the lowest energy domain wall switches type is $w = 61$~nm, which matches well with experimental results in Ref.~\cite{Koyama} where an equilibrium width between 40 and 76~nm was measured.} 
\end{center}
\end{figure}

A nanowire with PMA and two antiparallel domains can generate either Bloch or N\'eel domain wall configurations depending on the specific aspect ratio. Figure~\ref{energyswitch} illustrates this and compares the total energy (anisotropy, exchange, and demagnetizing) per unit area of Bloch (dashed line) and N\'eel walls (solid lines) at a constant nanowire thickness of $d=3.4$ nm as the width of the nanowire is increased. The integration range used to calculate the energy per unit area is between $z=\pm30$~nm (ie. $D=30$~nm), in order to include the domain wall and a portion of the domains. The particular material parameters, $M_{s} = 6.6\times10^{5}$~A/m, $A=15\times10^{-12}$ J/m, and $K=4.1\times10^{5}$ J/m$^{3}$, used to generate Fig.~\ref{energyswitch} were chosen as means of comparison to experimental findings by Koyama \emph{et al.} \cite{Koyama} The exchange constant was not included in the published work by Koyama, however, an average exchange constant for the layered ferromagnetic material comprising Co and Ni is assumed. 

In Fig.~\ref{numericcompare}(b) and Fig.~\ref{energyswitch} it can be seen that the total energy per unit area of the N\'eel wall decreases as the strip width decreases. This is due to the competition between anisotropy, exchange and demagnetizing energies. As the strip width decreases, so does the characteristic domain wall width $L_{N}$. This in turn means that the exchange energy increases. This energy increase is off-set by a larger decrease in the uniaxial anisotropy energy and also a decrease in the demagnetizing energy.

On the other hand, the energy per unit area of the Bloch wall increases gradually as the strip width is reduced, peaking at roughly $w=40$~nm in Fig.~\ref{energyswitch} and then decreasing. Again this is due to the competition between energies. If the dipolar energy per unit area is calculated using the expressions derived in the first iteration, then the Bloch wall energy monotonically decreases with decreasing strip width. However, in the second iteration there is a small peak that develops, mainly due to an increase in the demagnetizing energy produced by magnetostatic charges at the top and bottom surfaces of the strip. Equation~\eqref{demag2} does not adequately treat these surface charges as it assumes uniform magnetization in the calculation of demagnetizing factors and so the second iteration represents a more realistic calculation.

The difference between the slopes of the energies results in a crossing at strip width $w=61$~nm for the parameters of Fig.~\ref{energyswitch}. At this precise geometry, it is predicted that both Bloch and N\'eel domain wall types are stable and that domain walls may move with the least amount of energy.\cite{Jung} In particular, we find that it is the volume contribution to the dipolar energy per unit area of the N\'eel wall ($\frac{1}{2} \mu_{0} M_{s}^{2} N_{zz}$) that is most important in determining this crossing. Without it, the energies never cross. With it (even if the other dipolar energy terms are approximated using demagnetizing factors) a crossing is always seen. It is little wonder, therefore, that the first iteration of the theory produces no energy crossing as there is no mechanism for including dipolar energy due to magnetization gradients under the uniform magnetization approximation.

\begin{figure}[b]
\begin{center}
\includegraphics[width=8.5cm]{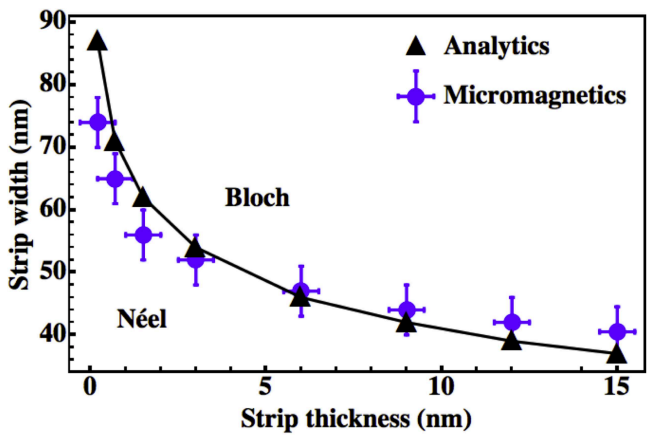}
\caption{\label{micromagneticcompare}(Color online) The two regions where Bloch and N\'eel walls represent the lowest energy domain walls, as a function of strip thickness and strip width. Equilibrium Bloch configurations are found above the solid line and N\'eel walls have lower energy below the solid line. The switch from one wall type to the other as calculated by Martinez \emph{et al.} \cite{Martinez} is shown by circles with error bars and the results generated by the analytic energy expressions developed in this work are shown by the triangles connected with the solid line. Material parameters used are $M_{s}=3\times10^{5}$ A/m, $A = 10\times10^{-12}$ J/m, and $K = 2\times10^{5}$ J/m$^{3}$. } 
\end{center}
\end{figure}

In the experimental work by Koyama \emph{et al.} \cite{Koyama}, it was found using resistivity measurements on nanowires with PMA, that both domain wall types occur in zero applied field at a nanowire width of $w=59$~nm. It was also found that when $w=40$~nm or $w=76$~nm (the closest widths measured either side of 59~nm) only one wall type or the other occurred. As can be seen in Fig.~\ref{energyswitch} the analytic switch for these parameters is calculated to occur at a nanowire width of $w=61$ nm, which agrees well with experimental findings. This is particularly remarkable given that ours is a one-dimensional model where approximations have been made for the domain wall width and also for the demagnetizing energy.

Since the range of widths measured in Ref.~\cite{Koyama} is limited, we also want to compare our analytic theory to other published works. Martinez \emph{et al.} \cite{Martinez} used micromagnetics and created a phase diagram for Bloch and N\'eel domain walls in thin ferromagnetic strips for a variety of aspect ratios using the material parameters for CoPtCr, namely $M_{s} = 3\times10^{5}$ A/m, $A = 10\times10^{-12}$ J/m, and $K = 2\times10^{5}$ J/m$^{3}$. These parameters were used in our Figs.~\ref{magapprox}-\ref{numericcompare} and are used in Fig.~\ref{micromagneticcompare}. In particular, in Fig.~\ref{numericcompare}(b) one can see an example of the switch between Bloch and N\'eel walls. The micromagnetic calculations of Martinez \emph{et al.} allowed the domain walls to have structure in both the $z$ and $x$ directions, as opposed to our assumption that the walls are 1D objects. 
 
Figure~\ref{micromagneticcompare} shows the exact numeric results of Martinez \emph{et al.} (circles with error bars) overlaid with our analytic results (triangles connected with solid lines) for the aspect ratios where Bloch and N\'eel walls exhibit equal energies. The error bars represent our errors in reading the results from Ref.~\cite{Martinez}. The solid line is a guide to the eye joining our analytic results. N\'eel walls occur for strip width (vertical axis) and thickness (horizontal axis) combinations below the solid line and Bloch walls have lower energy for width and thickness combinations above the line. As one can see, our values are in agreement within the micromagnetic error bars for most aspect ratios. 

The main discrepancy between the micromagnetic and analytic results in Fig.~\ref{micromagneticcompare} corresponds to the smallest nanowire thickness close to $d=0$. As $d\to0$, there is expected to be a large deviation in the surface demagnetizing energy of the system. This means that any tiny change in the value of $d$ that we use to obtain this point, compared to the value of $d$ that Martinez \emph{et al.} have used is expected to lead to very different results. We therefore feel that this discrepancy is of no considerable concern and our 1D analytic results appear robust. We have also compared the wall widths calculated using micromagnetics in Ref.~\cite{Martinez} (inferred from domain wall mobilities) to our analytic wall widths and the trends match well with less than a $2.5\%$ difference for each data point available for comparison.
 
 
\section{Conclusion}
\label{conclusion}
 
Our analytic approach to finding the switching point in energies between Bloch and N\'eel domain walls in a rectangular nanowire with PMA involves a two step iteration to approximate the magnetostatic energy. Comparison of results of this analytic theory with micromagnetic and experimental findings illustrate the success of our model. 

The analytic theory is simple to implement -- all the equations needed are provided in the main text and in the Appendix. It replaces the need for much more complicated and time-consuming calculations in order to predict the best geometry in order for nanowires to be at their Bloch-N\'eel switching point. It is therefore hoped to be of great use to scientists and engineers making domain wall devices.

A future extension of this work would be to add in additional energy contributions, such as the Dzyaloshinskii-Moriya interaction, to determine the change in the geometry for which the Bloch to N\'eel switch takes place. Also, our theory is only for static domain walls and the switching point may be altered when the domain wall is in motion under the force of applied fields or currents.


\begin{acknowledgments}

The authors thank P.~J.~Metaxas for useful discussions and the UCCS Letters Arts and Sciences Student-Faculty Research Award for support. M.D.D. acknowledges support from the UCCS Undergraduate Research Academy. 

 \end{acknowledgments}
 
 
\appendix*
\section{Auxiliary field}
\label{appendix}

Using Eqs.~\eqref{piecewise2}, \eqref{potential} and \eqref{strayfield}, the auxiliary field $\vec{H}(x,y,z)$ can be found and then averaged through the rectangular cross-section to find a one-dimensional dependence $\vec{H}(z)$.

The $x$ component of the auxiliary field applies only to the Bloch domain wall and is approximated by averaging the field over the $x$ and $y$ dimensions of the nanowire, as explained in the main text with reference to Fig.~\ref{Hfield}. $H_{x}$ is found to be

\begin{widetext}
\begin{eqnarray}
\label{Hxappendix}
\frac{4\pi LH_{x}(z)}{M_{s}} &=& \frac{d}{w} \sqrt{\frac{d^{2}}{4}+(z-L)^{2}} 
	- \frac{d}{w} \sqrt{\frac{d^{2}}{4}+w^{2}+(z-L)^{2}} 
	+\frac{d}{w} \sqrt{\frac{d^{2}}{4}+(z+L)^{2}} 
	- \frac{d}{w} \sqrt{\frac{d^{2}}{4}+w^{2}+(z+L)^{2}} \nonumber \\
&& -4(z-L) \arctan \left( \frac{d(z-L)}{2w\sqrt{\frac{d^{2}}{4}+w^{2}+(z-L)^{2}}} \right) 
	 -4(z+L) \arctan \left( \frac{d(z+L)}{2w\sqrt{\frac{d^{2}}{4}+w^{2}+(z+L)^{2}}} \right) \nonumber \\
&&  -\frac{2d}{w} \sqrt{\frac{d^{2}}{4}+z^{2}}
	+\frac{2d}{w} \sqrt{\frac{d^{2}}{4}+w^{2}+z^{2}}
	+8z \arctan \left( \frac{dz}{2w\sqrt{\frac{d^{2}}{4}+w^{2}+z^{2}}} \right)
	+\frac{4z^{2}}{w} \arctanh \left( \frac{2\sqrt{\frac{d^{2}}{4}+z^{2}}}{d} \right) \nonumber \\
&&+\frac{4dz}{w} \arctanh \left( \frac{\sqrt{\frac{d^{2}}{4}+z^{2}}}{z} \right)
	+\frac{4(w^{2}-z^{2})}{w} \arctanh \left( \frac{2\sqrt{\frac{d^2}{4}+w^{2}+z^{2}}}{d} \right)
	-\frac{4dz}{w} \arctanh \left( \frac{\sqrt{\frac{d^{2}}{4}+w^{2}+z^{2}}}{z} \right) \nonumber \\
&& -\frac{2(z-L)^{2}}{w} \arctanh \left( \frac{2\sqrt{\frac{d^{2}}{4}+(z-L)^{2}}}{d} \right)
	-\frac{2d(z-L)}{w} \arctanh \left( \frac{\sqrt{\frac{d^{2}}{4}+(z-L)^{2}}}{z-L} \right) \nonumber \\
&& -\frac{2(w^{2}-(z-L)^{2})}{w} \arctanh \left( \frac{2\sqrt{\frac{d^{2}}{4}+w^{2}+(z-L)^{2}}}{d} \right)
 	+\frac{2d(z-L)}{w} \arctanh \left( \frac{\sqrt{\frac{d^{2}}{4}+w^{2}+(z-L)^{2}}}{z-L} \right) \nonumber \\
&& -\frac{2(z+L)^{2}}{w} \arctanh \left( \frac{2\sqrt{\frac{d^{2}}{4}+(z+L)^{2}}}{d} \right)
	-\frac{2d(z+L)}{w} \arctanh \left( \frac{\sqrt{\frac{d^{2}}{4}+(z+L)^{2}}}{z+L} \right) \nonumber \\
&& -\frac{2(w^{2}-(z+L)^{2})}{w} \arctanh \left( \frac{2\sqrt{\frac{d^{2}}{4}+w^{2}+(z+L)^{2}}}{d} \right)
	+\frac{2d(z+L)}{w} \arctanh \left( \frac{\sqrt{\frac{d^{2}}{4}+w^{2}+(z+L)^{2}}}{z+L} \right), \nonumber \\
\end{eqnarray}
\end{widetext}
with $L=\pi L_{B}$. In the thin film limit ($w\to \infty$) the $x$ surfaces of the nanowire extend to infinity and the $H_{x}$ component vanishes.

For the $H_{y}$ component that is nonzero for both Bloch and N\'eel walls, the average value is approximated well by the value at the center of the strip $H_{y}(0,0,z)$, as explained in the main text. We therefore find that
\begin{widetext}
\begin{eqnarray}
\label{Hyappendix}
\frac{2\pi L H_{y}(z)}{M_{s}} &=& d \arccoth \left( \frac{w}{\sqrt{d^2 + w^2 +4 (L-z)^2}} \right) - d \arccoth \left(  \frac{w}{\sqrt{d^2 + w^2 +4 (L+z)^2}} \right)
\nonumber \\
&& + 2 (L-z) \arctan \left( \frac{2 w (L-z)}{ d \sqrt{d^2 + w^2 +4 (L-z)^2}} \right) - 2 (L+z) \arctan \left( \frac{2 w (L+z)}{ d \sqrt{d^2 + w^2 +4 (L+z)^2}} \right),
\end{eqnarray}
\end{widetext}
where $L= \ln(4) L_{N}$ for the N\'eel wall and $L=\ln(4) L_{B}$ for the Bloch wall. This is an odd function about $z=0$, as is expected. It also has the correct limit $|H_{y}| \to N_{y} M_{s}$ as $z\to \pm \infty$, to within a maximum 10\% error. However, in the range $d \ll w$ that we consider, the difference is less than 5\% and is due to the approximations made in calculating the auxiliary field. 

Similarly, the $z$ component  of the auxiliary field that is only non-zero for the N\'eel wall can be well-approximated by its value at the center of the strip $H_{z}(0,0,z)$. The expression we find is
\begin{widetext}
\begin{eqnarray}
\label{Hzappendix}
\frac{4\pi L H_{z}(z)}{M_{s}} &=& -4 (L-z) \arctan \left( \frac{d w}{2 (L-z) \sqrt{d^2 + w^2 +4 (L-z)^2}} \right) + 8 z \arctan \left( \frac{d w}{2 z \sqrt{d^2 + w^2 +4 z^2}}  \right) 
\nonumber \\
&&- 4 (L+z) \arctan  \left( \frac{d w}{2 (L+z) \sqrt{d^2 + w^2 +4 (L+z)^2}} \right) + w ~f(w,d,z) + d ~f(d,w,z),
\end{eqnarray}
where the function $f(w,d)$ is given by
\begin{eqnarray}
f(w,d,z) &=& - \ln \left( -d + \sqrt{d^2 + w^2 +4 (L-z)^2} \right) + \ln  \left( d + \sqrt{d^2 + w^2 +4 (L-z)^2} \right) 
\nonumber \\
&&+ 2 \ln \left( -d + \sqrt{d^2 + w^2 +4 z^2} \right) -2 \ln \left( d + \sqrt{d^2 + w^2 +4 z^2} \right)
\nonumber \\
&& - \ln \left( -d + \sqrt{d^2 + w^2 +4 (L+z)^2} \right) + \ln  \left( d + \sqrt{d^2 + w^2 +4 (L+z)^2} \right),
\end{eqnarray}
\end{widetext}
where $L= \pi L_{N}$. Notice that in the limit that the domain wall width $L\to 0$, then $H_{z}$ vanishes too.

The auxiliary field components $H_{x}(z)$, $H_{y}(z)$ and $H_{z}(z)$ (Eqs.~\eqref{Hxappendix},~\eqref{Hyappendix},~and~\eqref{Hzappendix}, respectively) and the linear magnetization approximations for $M_{x}(z)$, $M_{y}(z)$ and $M_{z}(z)$ (Eqs.~\eqref{piecewise}~and~\eqref{piecewise2}, along with the appropriate form for $M_{z}(z)$) can be substituted into Eq.~\eqref{demagenergy} to calculate the demagnetizing energy density for the Bloch and N\'eel domain walls as a function of distance from the domain wall center $z$. 

This energy density can be then be integrated over the range $z=\pm~D$ to give the demagnetizing energy per unit area as shown generally in Eq.~\eqref{demagenergy2}. The integrations are long. Many can be done by hand and some may be done using a symbolic package such as \emph{Mathematica}. In Eq.~\eqref{demagenergy2}, the $N_{xx}$ and $N_{yy}$ factors combine to give the demagnetizing energy of the Bloch wall and the $N_{yy}$ and $N_{zz}$ factors are added to describe the demagnetizing energy of the N\'eel configuration. The calculated $N_{xx}$, $N_{yy}$, and $N_{zz}$ factors are essential to describe the demagnetizing energy per unit area of the two wall types and are listed below.

\begin{widetext}
\begin{eqnarray}
 N_{xx} & = & -\frac{1}{288\pi w b_{xB}^{2}}
	\left[
		3d^{4} + \left( 104 d b_{xB}^{2} - 4 d^{3} \right) \sqrt{4b_{xB}^{2} + d^{2}} + \left(d^{3}-104db_{xB}^{2}\right) 	
					\sqrt{16b_{xB}^{2} + d^{2}} - \left(3d^{3}+30dw^{2}\right) 	
					\sqrt{d^{2}+4w^{2}} \right. \nonumber \\
		& & \left. + \left( 40dw^{2} + 4d^{3} -104db_{xB}^{2} \right) \sqrt{4b_{xB}^{2} + d^{2}+4w^{2}} + \left(104db_{xB}^{2}-
					d^{3}-10dw^{2} \right) \sqrt{16b_{xB}^{2} + d^{2}+4w^{2}} \right. \nonumber \\
		& & \left. +72w^{4} \arccoth\left({\frac{d}{\sqrt{d^{2}+4w^{2}}}}\right) + 6b_{xB}d^{3} \ln\left({d^{2}}\right) - 24b_{xB}
					d^{3} \ln\left({\sqrt{4b_{xB}^{2}+d^{2}}-2b_{xB}}\right) \right. \nonumber \\
		& & \left. +12b_{xB}d^{3} \ln\left({\sqrt{16b_{xB}^{2}+d^{2}}-4b_{xB}}\right) -6b_{xB}d^{3}\ln\left({d^{2}+4w^{2}}\right) 
					-12b_{xB}d^{3} \ln\left({\sqrt{16b_{xB}^{2}+d^{2}+4w^{2}}-4b_{xB}}\right) \right. \nonumber \\
		& & \left.  +24b_{xB}d^{3} \ln\left({\sqrt{4b_{xB}^{2}+d^{2}+4w^{2}}-2b_{xB}}\right) 
					- 24 \left(w^{4}-24w^{2}b_{xB}^{2}\right) \arccoth \left({\frac{d}{\sqrt{16b_{xB}^{2}+d^{2}+4w^{2}}}}
					\right) \right. \nonumber \\
		& & \left. - 48w^{2} \left(12b_{xB}^{2}+w^{2}\right) \arccoth \left({\frac{d}{\sqrt{4b_{xB}^{2}+d^{2}+4w^{2}}}}\right) 
					+192wb_{xB}^{3} \arctan \left({\frac{db_{xB}}{2w\sqrt{b_{xB}^{2}+d^{2}+w^{2}}}}\right) \right. \nonumber \\
		& & \left. +\left(768wb_{xB}^{3}-192w^{3}b_{xB}\right) \arctan \left({\frac{2db_{xB}}{w\sqrt{16b_{xB}^{2}+d^{2}+4w^{2}}}}
					\right) \right. \nonumber \\
		& & \left. +\left(384w^{3}b_{xB}-576wb_{xB}^{3}\right) \arctan \left({\frac{db_{xB}}{w\sqrt{4b_{xB}^{2}+d^{2}+4w^{2}}}}
					\right) -192b_{xB}^{4} \ln\left({\frac{\sqrt{16b_{xB}^{2} +d^{2}}-d}{\sqrt{16b_{xB}^{2}+d^{2}+4w^{2}}-d}}
					\right) \right. \nonumber \\
		& & \left. +192b_{xB}^{4} \ln\left({\frac{\sqrt{16b_{xB}^{2} +d^{2}}+d}{\sqrt{16b_{xB}^{2}+d^{2}+4w^{2}}+d}}
					\right)  +48b_{xB}^{4} \ln\left({\frac{d-\sqrt{4b_{xB}^{2} +d^{2}}}{d-\sqrt{4b_{xB}^{2}+d^{2}+4w^{2}}}}
					\right) \right. \nonumber \\
		& & \left. +72w^{4} \ln\left({\frac{d-\sqrt{d^{2}+4w^{2}}}{d-\sqrt{4b_{xB}^{2}+d^{2}+4w^{2}}}}
					\right)-48b_{xB}^{4} \ln\left({\frac{\sqrt{4b_{xB}^{2} +d^{2}}+d}{\sqrt{4b_{xB}^{2}+d^{2}+4w^{2}}+d}}
					\right) \right. \nonumber \\
		& & \left. -72w^{4} \ln\left({\frac{\sqrt{d^{2}+4w^{2}}+d}{\sqrt{4b_{xB}^{2}+d^{2}+4w^{2}}+d}}
					\right) 
						-24b_{xB}d\left(
						2b_{xB}^{2} \ln\left({d^{2}}\right) - 8b_{xB}^{2} \ln\left({\sqrt{4b_{xB}^{2}+d^{2}}-2b_{xB}}\right) 	
								\right. \right. \nonumber \\
						& & \left. \left. - 4b_{xB}^{2} \ln\left({\sqrt{4b_{xB}^{2}+d^{2}}+2b_{xB}}\right)+ 8b_{xB}^{2} \ln
								\left({\sqrt{16b_{xB}^{2}+d^{2}}-4b_{xB}}\right) +\left(3w^{2}-2b_{xB}^{2}\right) \ln
								\left({d^{2}+4w^{2}}\right) \right. \right. \nonumber \\
						& & \left. \left. +\left(6w^{2}-8b_{xB}^{2}\right) \ln\left({\sqrt{16b_{xB}^{2}+d^{2}+4w^{2}}-4b_{xB}}
								\right) \right. \right. \nonumber \\
						& & \left. \left. +4b_{xB}^{2} \ln\left({\frac{\left(\sqrt{16b_{xB}^{2} + d^{2}}-4b_{xB}\right)
								\left(\sqrt{16b_{xB}^{2} + d^{2}+4w^{2}}+4b_{xB}\right)}{\left(\sqrt{16b_{xB}^{2} + d^{2}}
								+4b_{xB}\right)\left(\sqrt{16b_{xB}^{2} + d^{2}+4w^{2}}-4b_{xB}\right)}}\right) \right. 
								\right. \nonumber \\
						& & \left. \left. +\left(8b_{xB}^{2}-12w^{2}\right)\ln\left({\sqrt{4b_{xB}^{2}+d^{2}+4w^{2}}-2b_{xB}}
								\right) +4b_{xB}^{2} \ln\left({\sqrt{4b_{xB}^{2}+d^{2}+4w^{2}}+2b_{xB}}\right) \right. 
								\right. \nonumber \\
						& & \left. \left. +b_{xB}^{2} \ln\left({\frac{\left(d^{2}+8b_{xB}^{2}+4b_{xB}\sqrt{4b_{xB}^{2}+d^{2}}\right)\left(d^{2}+8b_{xB}^{2}+4w^{2}-4b_{xB}\sqrt{4b_{xB}^{2}+d^{2}+4w^{2}}\right)}{\left(d^{2}+8b_{xB}^{2}-4b_{xB}\sqrt{4b_{xB}^{2}+d^{2}}\right)\left(d^{2}+8b_{xB}^{2}+4w^{2}+4b_{xB}\sqrt{4b_{xB}^{2}+d^{2}+4w^{2}}\right)}}\right) 
						\right)
	\right],
\label{Nxx}
\end{eqnarray}
Notice that this demagnetizing energy per unit area factor only depends on the nanowire geometry ($d$ and $w$) and on the adjustable fit length $b_{xB}$ for the Bloch wall width.


\begin{eqnarray}
 N_{yy} & = & -\frac{1}{24\pi b_{y}^{2}}\left[
 		dw\sqrt{16b_{y}^{2}+d^{2}+w^{2}}-dw\sqrt{d^{2}+w^{2}}+\left(12d^{2}b_{y}-64b_{y}^{3}\right)\arctan \left({\frac{4 w b_{y}}
				{d\sqrt{16b_{y}^{2}+d^{2}+w^{2}}}}\right) \right. \nonumber \\
		& & \left. +\left(6d^{2}b_{y}+48Db_{y}^{2}-24b_{y}^{3}-24D^{2}b_{y}\right)\arctan \left({\frac{2 w (D-b_{y})}
				{d\sqrt{4\left(D-b_{y}\right)^{2}+d^{2}+w^{2}}}}\right) \right. \nonumber \\
		& & \left. +\left(48Db_{y}^{2}-6d^{2}b_{y}+24b_{y}^{3}+24D^{2}b_{y}\right)\arctan \left({\frac{2 w (D+b_{y})}
				{d\sqrt{4\left(D+b_{y}\right)^{2}+d^{2}+w^{2}}}}\right) -d^{3}\arctanh\left({\frac{\sqrt{d^{2}+w^{2}}}{w}}\right)
				\right. \nonumber \\
		& & \left. +d^{3}\arctanh\left({\frac{\sqrt{16b_{y}^{2}+d^{2}+w^{2}}}{w}}\right) +3wdb_{y}\ln\left({4}\right)+6wdb_{y}\left(\ln
				\left({2}\right)-2\ln\left({4b_{y}+\sqrt{16b_{y}^{2}+d^{2}+w^{2}}}\right)\right) \right. \nonumber \\
		& & \left. -24db_{y}^{2}\ln\left({1+\frac{\sqrt{16b_{y}^{2}+d^{2}+w^{2}}}{w}}\right)+24db_{y}^{2}\ln\left({1-
				\frac{\sqrt{16b_{y}^{2}+d^{2}+w^{2}}}{w}}\right) \right. \nonumber \\
		& & \left. +6wdb_{y} \ln\left({b_{y}-D+\frac{1}{2}\sqrt{4\left(D-b_{y}\right)^{2}+d^{2}+w^{2}}}\right)
				+\left(12db_{y}^{2}-12dDb_{y}\right) \ln\left({1+\frac{\sqrt{4\left(D-b_{y}\right)^{2}+d^{2}+w^{2}}}{w}}\right) 
				\right. \nonumber \\
		& & \left. +\left(12dDb_{y}-12db_{y}^{2}\right) \ln\left({1-\frac{\sqrt{4\left(D-b_{y}\right)^{2}+d^{2}+w^{2}}}{w}}\right) 
				+6wdb_{y} \ln\left({b_{y}+D+\frac{1}{2}\sqrt{4\left(D+b_{y}\right)^{2}+d^{2}+w^{2}}}\right) \right. \nonumber \\
		& & \left. +\left(12dDb_{y}+12db_{y}^{2}\right) \ln\left({1+\frac{\sqrt{4\left(D+b_{y}\right)^{2}+d^{2}+w^{2}}}{w}}\right) 
				\right. \nonumber \\
		& & \left. -\left(12dDb_{y}+12db_{y}^{2}\right) \ln\left({1-\frac{\sqrt{4\left(D+b_{y}\right)^{2}+d^{2}+w^{2}}}{w}}\right) 
 \right],
 \label{Nyy}
\end{eqnarray}
This demagnetizing energy per unit area factor depends on the nanowire geometry ($d$ and $w$), on the adjustable fit length ($b_{yB}$ for the Bloch wall width or $b_{yN}$ for the N\'eel wall, written here as just $b_y$), and on the integration range $D$ since the domains have a non-zero contribution to this part of the energy per unit area.


\begin{eqnarray}
 N_{zz} & = & -\frac{1}{48 \pi b_{zN}^{2}} 
  \left[
	12dw \sqrt{d^{2} + w^{2}} 
	- 16dw \sqrt{4 b_{zN}^{2} + d^{2} + w^{2}}
	+ 4dw \sqrt{16  b_{zN}^{2} + d^{2} + w^{2}} 
	- 12w^{3}  \arccoth\left({\frac{d}{\sqrt{d^{2}+w^{2}}}} \right) \right. \nonumber \\
	& & \left. + 16w^{3} \arccoth \left({\frac{d}{\sqrt{4b_{zN}^{2}+d^{2}+w^{2}}}} \right) 
	 - 4w^{3} \arccoth \left({\frac{d}{\sqrt{16b_{zN}^{2}+d^{2}+w^{2}}}} \right) \right. \nonumber \\
	   & & \left. +16d^{3} \arccoth \left({\frac{w}{\sqrt{4b_{zN}^{2}+d^{2}+w^{2}}}} \right)	 
	   -12d^{3} \arccoth \left({\frac{w}{\sqrt{d^{2}+w^{2}}}} \right)
	  -4d^{3} \arccoth \left({\frac{w}{\sqrt{16b_{zN}^{2}+d^{2}+w^{2}}}} \right) \right. \nonumber \\
	& & \left.  -48w^{2}b_{zN} \arctan \left({\frac{2db_{zN}}{w\sqrt {4b_{zN}^{2}+d^{2}+w^{2}}}} \right) 
	  -64b_{zN}^{3} \arctan \left({\frac{dw}{2b_{zN}\sqrt {4b_{zN}^{2}+d^{2}+w^{2}}}} \right)  \right. \nonumber \\
	 & & \left. +24w^{2}b_{zN} \arctan \left({\frac{4db_{zN}}{w\sqrt {16b_{zN}^{2}+d^{2}+w^{2}}}} \right)
	   +24d^{2}b_{zN} \arctan \left({\frac{4wb_{zN}}{d\sqrt {16b_{zN}^{2}+d^{2}+w^{2}}}} \right) \right. \nonumber \\
	   & & \left.  - 48d^{2}b_{zN} \arctan \left({\frac{2wb_{zN}}{d\sqrt {4b_{zN}^{2}+d^{2}+w^{2}}}} \right) 
	+128b_{zN}^{3} \arctan \left(\frac{dw}{4b_{zN}\sqrt{16b_{zN}^{2}+d^{2}+w^{2}}} \right)  \right. \nonumber \\
	 & & \left. -9w^{3} \ln \left({\sqrt{d^{2}+w^{2}}-d}\right)
	 +9w^{3} \ln \left({\sqrt{d^{2}+w^{2}}+d}\right) 
	  + \left(12w^{3}-48wb_{zN}^{2}\right) \ln \left({\sqrt{4b_{zN}^{2}+d^{2}+w^{2}}-d}\right)\right. \nonumber \\
	   & & \left.  +\left(48wb_{zN}^{2}-12w^{3}\right) \ln \left({\sqrt{4b_{zN}^{2}+d^{2}+w^{2}}+d}\right) 
	     +\left(48wb_{zN}^{2}-3w^{3}\right) \ln \left({\sqrt{16b_{zN}^{2}+d^{2}+w^{2}}-d}\right) \right. \nonumber \\
	    & & \left. +\left(3w^{3} -48wb_{zN}^{2}\right) \ln \left({\sqrt{16b_{zN}^{2}+d^{2}+w^{2}}+d}\right)
	 -9d^{3} \ln \left({\sqrt{d^{2}+w^{2}}-w} \right)
	  +9d^{3} \ln \left({\sqrt{d^{2}+w^{2}}+w} \right)  
	  \right. \nonumber \\
	  & & \left.
	   +12d^{3} \ln \left({\sqrt{4b_{zN}^{2}+d^{2}+w^{2}}-w} \right) 
	   -12d^{3} \ln \left({\sqrt{4b_{zN}^{2}+d^{2}+w^{2}}+w} \right)
	  -3d^{3} \ln \left({\sqrt{16b_{zN}^{2}+d^{2}+w^{2}}-w} \right) 
	   \right. \nonumber \\
	  & & \left.
	   +3d^{3} \ln \left({\sqrt{16b_{zN}^{2}+d^{2}+w^{2}}+w} \right) 
	    -24wdb_{zN}\ln\left({d^{2}+w^{2}} \right)
	  +96wdb_{zN}\ln\left({\sqrt{4b_{zN}^{2}+d^{2}+w^{2}}+2b_{zN}}\right) 
	   \right. \nonumber \\
	  & & \left.
	    -48wdb_{zN}\ln\left({\sqrt{16b_{zN}^{2}+d^{2}+w^{2}}+4b_{zN}}\right)
	  -48db_{zN}^{2}\ln\left({\sqrt{4b_{zN}^{2}+d^{2}+w^{2}}-w}\right)  \right. \nonumber \\
	  & & \left. +48db_{zN}^{2}\ln\left({\sqrt{4b_{zN}^{2}+d^{2}+w^{2}}+w}\right)
	   +48db_{zN}^{2}\ln\left({\sqrt{16b_{zN}^{2}+d^{2}+w^{2}}-w}\right) \right. \nonumber \\ 
	  & & \left. -48db_{zN}^{2}\ln\left({\sqrt{16b_{zN}^{2}+d^{2}+w^{2}}+w}\right)
  \right].
  \label{Nzz}
\end{eqnarray}

Again, we have dependence on the nanowire geometry ($d$ and $w$) and this time on the adjustable fit length $b_{zN}$ for the N\'eel wall width.
\end{widetext}


\end{document}